\documentclass[aps,prl,twocolumn,superscriptaddress]{revtex4}
\usepackage{amssymb,amsmath}
\usepackage{graphicx}
\usepackage{hyperref}

\begin{document}

\title{Percolation quantum phase transitions in diluted magnets}
\author{Thomas Vojta}
\affiliation{Department of Physics, University of Missouri-Rolla, Rolla, MO 65409}
\author{J\"org Schmalian}
\affiliation{Department of Physics and Astronomy and Ames Laboratory, Iowa State
University, Ames, IA 50011}
\date{\today}

\begin{abstract}
We show that the interplay of geometric criticality and quantum fluctuations leads to a
novel universality class for the percolation quantum phase transition in diluted magnets.
All critical exponents involving dynamical correlations are different from the classical
percolation values, but in two dimensions they can nonetheless be determined exactly. We
develop a complete scaling theory of this transition, and we relate it to recent
experiments in La$_{2}$Cu$_{1-p}$(Zn,Mg)$_{p}$O$_{4}$. Our results are also relevant for
disordered interacting boson systems.

\end{abstract}

\maketitle


%%%%%%%%%%%%%%%%%%%%%%%%%%%%%%%%%%%%%%%%%%%%%%%%%%%%%%%%%%%%%%%%%%%%%%%%%%%%%%%%%%%%%%%%%

Geometric criticality and quantum criticality are two distinct phenomena leading to
universal scale invariant correlations. Both get combined in randomly diluted quantum
magnets. Site or bond dilution defines a percolation problem for the lattice with a
geometric phase transition at the percolation threshold \cite{percolation}. Quantum
fluctuations of the spins coexist with geometric fluctuations due to percolation. In this
Letter, we address the question of whether or not these quantum fluctuations
fundamentally change the percolation phase transition.

\begin{figure}[tbp]
\includegraphics[width=7.8cm]{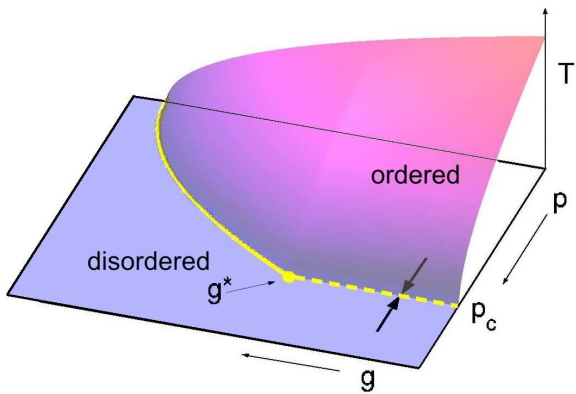}
\caption{Schematic phase diagram of a diluted magnet with impurity concentration $p$,
temperature $T$, and quantum fluctuation strength $g$ \cite{dim}.
There is a quantum multicritical point (big dot) at ($p_{c},g^{\ast }$%
).  The zero temperature quantum phase transition across the dashed line is the topic of
this Letter. } \label{fig:pd}
\end{figure}

The generic phase diagram of a diluted magnet as function of impurity concentration $p$,
quantum fluctuations strength $g$, and temperature $T$ is shown in Fig.\ \ref{fig:pd}.
In the absence of both thermal and quantum fluctuations ($g=T=0$), magnetic long-range
order survives for all concentrations $p\leq p_{c}$, where $p_{c}$ is the classical
(geometric) percolation threshold. If $p>p_{c}$, no long-range order is possible as the
system is decomposed into independent clusters.  For $p<p_{c}$, magnetic order survives
up to a nonzero temperature $T_{c}(p)$; but at  $p_{c}$, it is destroyed immediately by
thermal fluctuations, $T_c(p_c)=0$ \cite{Bergstresser,Stephen,Gefen,ramification}.

Zero temperature quantum fluctuations are less effective in destroying magnetic order.
Even at $p_{c}$, the percolating cluster can remain ordered up to a nonzero fluctuation
strength. This was found in the diluted transverse field Ising model
\cite{oldising,SenthilSachdev} where a multi-critical point with transverse field
strength $g^{\ast }$ and $p=p_c$ emerges. At this point, both quantum and geometric
fluctuations diverge.
In $O(N)$ systems ($N\ge 2$), early results suggested that quantum fluctuations always
destroy long-range order at $p_c$ \cite{Wan,YasudaChen,Kato}. However, Sandvik
\cite{Sandvik_AFM} showed that the percolating cluster in a 2d diluted Heisenberg quantum
antiferromagnet (QAFM) is ordered at $p_{c}$ (see also Ref.\ \cite{rotor_sw}). If quantum
fluctuations are enhanced, e.g., by an interlayer coupling in a bilayer QAFM
\cite{bilayerS,bilayerV}, a multi-critical point at ($p_{c},g^{\ast }$) arises, similar
to the transverse field Ising case.  The behavior at the generic quantum phase transition
occurring for $g>g^{\ast }$, $p<p_{c}$ (solid line in Fig.\ \ref{fig:pd}) and at the
multi-critical point at ($p_{c},g^{\ast }$) has been investigated in the past
\cite{bilayerS,bilayerV,us_hsb04}. In contrast, the percolation quantum phase transition
at $p_{c}$ and $g<g^{\ast }$ which can be observed, e.g.,
in diluted 2d QAFMs such as La$_{2}$Cu$_{1-p}$(Zn,Mg)%
$_{p}$O$_{4}$ \cite{Vajk_exp} has received less attention.

In this Letter, we show that the interplay of geometric criticality and quantum
fluctuations leads to a novel universality class for this percolation quantum phase
transition. Even though the transition is driven entirely by geometric criticality of the
underlying classical percolation problem, quantum fluctuations enhance the singularities
in all quantities involving dynamic correlations. As a result, the susceptibility
exponent $\gamma $ as well as $\alpha $, $\delta $ and $\eta $ differ from the classical
percolation values, and the dynamical exponent $z$ is nontrivial. In contrast, the order
parameter and correlation lengths exponents are classical, $\beta =\beta _{c}$ and $\nu
=\nu _{c}$ \cite{subscript}. In the remainder of this Letter we sketch the derivation of
these results via a scaling approach (we have also performed detailed explicit
calculations) and calculate critical exponent values in 2d and 3d. In
addition, we make predictions for various experimentally relevant observables.

Our starting point is a general $O(N)$ quantum rotor model ($N\ge 2$) on a
$d$-dimensional hypercubic lattice with nearest neighbor interactions as arises as
low-energy theory of a Heisenberg quantum antiferromagnet \cite{sachdevbook}. The action
reads
\begin{eqnarray}
\mathcal{A} &=&\int d\tau \sum_{\langle ij\rangle }J\epsilon _{i}\epsilon
_{j}\mathbf{S}_{i}(\tau )\cdot \mathbf{S}_{j}(\tau )+\sum_{i}\epsilon _{i}%
\mathcal{A}_{\mathrm{dyn}}[\mathbf{S}_{i}]  \notag \\
\mathcal{A}_{\mathrm{dyn}}[\mathbf{S}] &=&\frac{T}{g}\sum_{n=-\infty}^\infty |\omega
_{n}|^{2/z_{0}}\mathbf{S}(\omega _{n})\mathbf{S}(-\omega _{n})~. \label{eq:action}
\end{eqnarray}%
Here $\mathbf{S}_{i}(\tau )$ is a $N$-component unit vector at lattice site $i$ and
imaginary time $\tau$. The quenched random variable $\epsilon _{i}$ describes the site
dilution. It takes the values $0$ and $1$ with probabilities $p$ and $(1-p)$,
respectively. The parameter $g$ in the dynamic term measures the strength of the quantum
fluctuations \cite{fluc}, $\omega_n$ are bosonic Matsubara frequencies, and $z_{0}$ is
the dynamical exponent of the clean system. We are mostly interested in the case
$z_{0}=1$ (undamped rotor dynamics) but our results also apply for $0<z_{0}<2$ where
$z_{0}\rightarrow 0$ corresponds to the classical limit while $z_{0}=2$ is the case of
overdamped spin dynamics (occurring in itinerant magnets \cite{Hertz76,itfootnote}).

\begin{figure}[tbp]
\includegraphics[width=0.75\columnwidth]{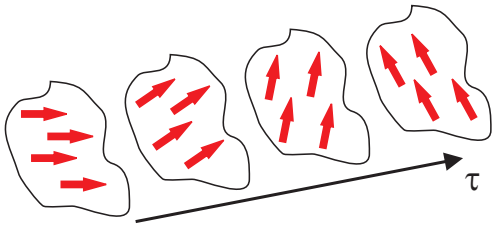}
\caption{For $g<g^{\ast }$, a percolation cluster acts as a compact object. The spins are
correlated in space but collectively fluctuate in imaginary time $\protect\tau $.}
\label{fig:dynamics}
\end{figure}

Let us start our discussion of the percolation quantum phase transition at $p=p_{c}$ and
$g<g^\ast$ by considering order parameter $m$ and correlation length $\xi$.  Long range
magnetic order can only develop on the infinite percolation cluster; all finite clusters
are completely decoupled. Since for $g<g^{\ast }$ the infinite cluster remains
magnetically ordered for all $p\leq p_{c}$, the total order parameter is proportional to
the number of sites in the
infinite cluster, $m\sim P_{\infty }(p)\sim (p_{c}-p)^{\beta _{c}}$. Thus, $%
\beta =\beta _{c}$. To determine the magnetic correlation length $\xi $ we note that the
correlations cannot extend beyond the connectedness length $\xi _{c}$ of the percolating
lattice because the rotors on different clusters are decoupled. On the other hand, for
$g<g^{\ast }$, all rotors on the same cluster are strongly correlated in space even
though they collectively fluctuate in time (see Fig.\ \ref{fig:dynamics}).
We thus conclude $\xi \sim \xi _{c}$ and $\nu =\nu _{c}$.

To study quantities involving dynamic correlations, we develop a scaling theory of
the free energy, contrasting classical and quantum cases. Consider a single percolation
cluster of finite size (number of sites) $s$. In  a classical magnet for
$T\rightarrow 0$, the free energy of such a cluster has the scaling form $%
F_{s}^{\mathrm{c}}\left( h,T\right) =T\Phi \left( hs/T\right) $, where $h$
is the field conjugate to the order parameter and $\Phi \left( x\right) $ a
dimensionless function. The susceptibility $\chi _{s}^{%
\mathrm{c}}$ follows from taking the second derivative with respect to $h$; this gives
$\chi _{s}^{\mathrm{c}}\sim s^{2}/T$ as expected. Let us now consider the quantum case.
For $g<g^{\ast }$, a quantum rotor cluster of size $s$ can be treated as a compact object
that fluctuates in time only (see Fig.\ \ref{fig:dynamics}) and is described by a 0+1
dimensional nonlinear sigma model (NLSM) with the action \cite{Fisher89,Toner}
\begin{equation}
\mathcal{A}_{s}=s\mathcal{A}_{\mathrm{dyn}}\left[ \mathbf{S}\right] +hs\int
d\tau S^{\left( 1\right) }\left( \tau \right) .  \label{eg:clusternlsm}
\end{equation}%
Here, $S^{\left( 1\right) }$ is one component of  $\mathbf{S}$. Dimensional analysis
shows that the free energy of a cluster of size $s$ is
\begin{equation}
F_{s}\left( g,h,T\right) =g^{\varphi }s^{-\varphi }\Phi \left( hs^{1+\varphi
}g^{-\varphi },Ts^{\varphi }g^{-\varphi }\right)  \label{eq:Fsq}
\end{equation}%
with exponent $\varphi =z_{0}/(2-z_{0})$ (for $z_{0}<2$). This result can be obtained by
rescaling imaginary time, $\tau \rightarrow g^{\varphi }s^{-\varphi }\tau $, which
eliminates the factor $(s/g)$ in the dynamic
part of the action, and leads to a dimensionless field strength $%
(hs^{1+\varphi }/g^{\varphi })$ and temperature $(Ts^{\varphi }/g^{\varphi }) $. The
resulting order parameter susceptibility  $\chi _{s}\sim s^{2+\varphi }$ of a quantum
spin cluster diverges more strongly with $s$ than that of a classical spin cluster. We
have obtained the same results from an explicit large-$N$ calculation \cite{Vojta05} and
from a renormalization group analysis of the NLSM (\ref%
{eg:clusternlsm}) at its strong coupling fixed point \cite%
{Toner,exact}

After having analyzed a single cluster we turn to the full percolation problem. The total
free energy can be obtained by summing the cluster free energy $F_s$ over all percolation
clusters. From classical percolation theory, we know that close to $p_{c}$ the number
$n_{s}$ of occupied clusters of size $s$ per lattice site obeys the scaling form
\begin{equation}
n_{s}\left( t\right) =s^{-\tau_c }f\left( ts^{\sigma_c }\right) .
\label{percscaling}
\end{equation}
Here $t=p-p_c$, and $\tau_c $ and $\sigma_c$ are classical percolation exponents. The
scaling function $f(x)$ is constant for small $x$ and drops off rapidly for large $x$.
All classical percolation exponents are
determined by $\tau_c$ and $\sigma_c$ including  $\nu_c =({\tau_c -1})/{%
(d\sigma_c )}$ and the fractal dimension $D_f=d/(\tau_c-1)$ of the
percolating cluster \cite{percolation}.

We first briefly recapitulate the percolation transition of a classical
magnet at $T=0$. The total free energy is given by the sum $F^{\mathrm{c}%
}\left( t,h,T\right) =\sum_{s}n_{s}\left( t\right) F_{s}^{\mathrm{c}}\left( h,T\right)$.
Rescaling $s\rightarrow s/b^{D_f}$ and taking the limit $T\to 0$ yields
\begin{equation}
F^{\mathrm{c}}\left( t,h\right) =b^{-d}F^{\mathrm{c}}\left( tb^{1/\nu_c
},hb^{D_f}\right)~.  \label{eq:scalin_class}
\end{equation}
The critical behavior of thermodynamic quantities can be obtained by taking the
appropriate derivatives. All critical exponents coincide with the classical (geometric)
percolation exponents which is expected because in a classical magnet at $T=0$ geometric
fluctuations are the only fluctuations
present \cite{percolation,Bergstresser,Gefen}. %The exponents can be
%expressed, e.g., in terms of $\nu_c$ and $D_f$:
%\begin{subequations}
%\begin{eqnarray}
%2-\alpha_c &=& d \nu_c \\
%\beta_c &=& \left( d-D_f\right) \nu_c, \\
%\gamma_c &=& \left( 2D_f-d\right) \nu_c , \\
%\delta_c &=& D_f/(d-D_f)
%\end{eqnarray}

We next generalize these ideas to the percolation quantum phase transition
in the quantum rotor model (\ref{eq:action}) for $g<g^{\ast }$. Averaging
the free energy (\ref{eq:Fsq}) of a quantum rotor cluster over all cluster
sizes, $F\left( t,h,T\right) =\sum_{s}n_{s}\left( t\right) F_{s}\left(
g,h,T\right) $, and rescaling $s\rightarrow s/b^{D_{f}}$ yields
\begin{equation}
F\left( t,h,T\right) =b^{-\left( d+z\right) }F\left( tb^{1/\nu },hb^{\left(
D_{f}+z\right) },Tb^{z}\right)  \label{eq:scaling}
\end{equation}%
with the correlation length exponent identical to the classical value, $\nu =\nu _{c}$.
$z=\varphi D_{f}$ plays the role of the dynamic critical exponent. We have verified this
interpretation by calculating the correlation time $\xi _{\tau }$ and showing $\xi _{\tau
}\sim \xi ^{z}$. Comparing the scaling form of the free energy (\ref%
{eq:scaling}) with its classical counterpart (\ref{eq:scalin_class}) we notice that the
quantum theory is obtained from the classical one, as usual, by replacing $d$ by $d+z$.
In addition, one should replace $D_{f}$ by $D_{f}+z$. This occurs since the magnetic
field couples to the spins at all "imaginary time points".

The critical exponents at the percolation quantum phase transition can be obtained by
taking the appropriate derivatives of the free energy. All exponents are fully determined
by two classical percolation exponents (e.g., the
fractal dimension $D_{f}$ and the correlation length exponent $\nu =\nu _{c}$%
) as well as the dynamic exponent $z$ (the classical exponents are recovered by setting
$z=0$):
\begin{eqnarray}
2-\alpha  &=&\left( d+z\right) \nu  \\
 \beta  &=&\left( d-D_{f}\right) \nu  \\
 \gamma &=&\left( 2D_{f}-d+z\right) \nu  \\
 \delta  &=&({D_{f}+z})/({d-D_{f}}) \\
2-\eta &=&  2D_{f}-d+z~.
\label{exponents}
\end{eqnarray}%
The exponents $\alpha $, $\gamma $, $\delta $, and $\eta$ are modified compared to classical
percolation, while $\beta $ depends only on $d-D_{f}$ and is thus unchanged. The
exponents fulfill the usual scaling laws, and hyperscaling is valid for $d<6$. This
means, the upper critical dimension is identical to its classical value, it is \emph{not} changed by the substitution $%
d\rightarrow d+z$.

The percolation exponents are known exactly in 2d \cite{percolation}, as is the dynamical
exponent $z=D_{f}\varphi =D_{f}z_{0}/(2-z_{0})$. Thus, all exponents for the
quantum phase transition can be determined exactly. For higher
dimensions, $\nu $ and $D_{f}$ are known numerically with good accuracy \cite%
{percolation}. Table I shows exponent values for $z_0=1$ in 2d and 3d.
The dependence of the critical exponents  on $z_{0}$ is shown in Fig.\ \ref%
{fig:exponents}.
In the limit $z_{0}\rightarrow 0$, the quantum exponents become identical to the
classical percolation exponents, and the dynamical exponent $z$ vanishes. For
$z_{0}\rightarrow 2$, the dynamical exponent $z$ as well as $\gamma $ and $%
\delta $ diverge, indicating activated scaling at an infinite-randomness critical point
as in the Ising case \cite{SenthilSachdev}.
%We will come back to this point below.
\begin{table}[tbp]
\begin{tabular}{c|cc|cc}
\hline & \multicolumn{2}{c}{2d} & \multicolumn{2}{c}{3d} \\
Exponent & classical & quantum & classical & quantum \\
\hline
$\alpha $ & $-$2/3 & $-$115/36 & $-$0.62 & $-$2.83 \\
$\beta $ & 5/36 & 5/36 & 0.417 & 0.417 \\
$\gamma $ & 43/18 & 59/12 & 1.79 & 4.02 \\
$\delta $ & 91/5 & 182/5 & 5.38 & 10.76 \\
$\nu $ & 4/3 & 4/3 & 0.875 & 0.875 \\
$\eta $ & 5/24 & $-$27/16 & $-$0.06 & $-$2.59 \\
$z$ & - & 91/48 & - & 2.53 \\ \hline
\end{tabular}%
\caption{Critical exponents at the classical and quantum percolation transition for $d=2$
and $3$ in case of  $z_{0}=1$. }
\end{table}
\begin{figure}[tbp]
\includegraphics[width=0.9\columnwidth,clip=true]{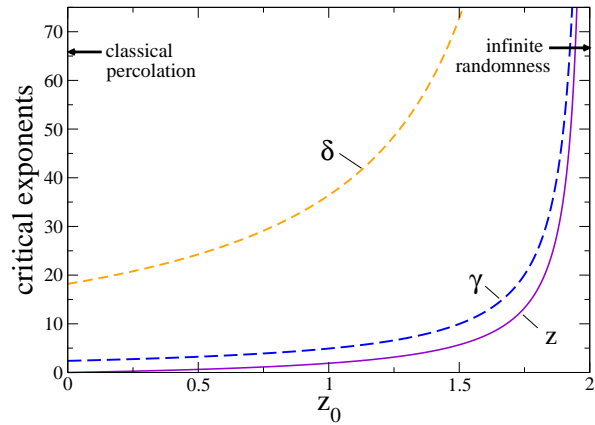}
\caption{Exponents $\gamma $, $\delta $, and $z$ as functions of $z_{0}$ in 2d. The
limiting cases $z_{0} \to 0$ and $z_{0} \to 2$ are explained in the text.}
\label{fig:exponents}
\end{figure}

The scaling form (\ref{eq:scaling}) of the free energy also determines the temperature
dependence of the order parameter susceptibility $\chi \left( t,T\right) =T^{-{\gamma
}/({\nu z})}\Theta _{\chi }\left( t^{\nu z}/T\right) $ and the heat capacity
$C=T({\partial ^{2}F}/{\partial T^{2}})=T^{d/z}\Theta _{c}\left( t^{\nu z}/{T}\right) $
with  scaling functions $\Theta _{\chi ,c}\left( x\right) $. At $p=p_{c}$ \ (or for
$p-p_{c}<T^{1/(\nu z)}$),
this yields $\chi \left( T\right) \propto T^{-\gamma /(\nu z)}$ and \ $%
C\left( T\right) \propto T^{d/z}$. Spin wave type excitations \emph{within} a cluster
have typical energies $\omega _{\mathrm{sw}}\propto s^{-z_{0}/D_f}$ that are large
compared to the energy gap $\omega _{0}\propto s^{-\varphi }=s^{-z/D_f}$ of the cluster
because $z>z_{0}$ (for $z_0=1$). Thus, spin waves that were shown to be important away
from $p_{c}$ \cite{Chernyshev02}, are subleading at $p_{c}$ as can also directly be seen
from the heat capacity contributions.

A scaling theory for the dynamic susceptibility can be derived following similar
arguments. We obtain $\chi ( t,\omega ,T) =b^{\gamma /\nu }\chi ( tb^{1/\nu },\omega
b^{z},Tb^{z})$ which at $p_c$ yields on the real frequency axis:
\begin{equation}
\mathrm{Im}\chi \left( \omega +i0^{+}\right) =\omega ^{-\gamma /(\nu
z)}\Omega \left( \omega /T\right) .  \label{eq:Imsus}
\end{equation}%
with dimensionless scaling function $\Omega \left( x\right)$.
%that can be calculated, e.g., in the large-$N$ limit of the quantum rotor model.
 Eq.\ (\ref{eq:Imsus}) implies that
low-temperature inelastic neutron scattering
experiments at the location of the Bragg peak should see a sharp increase $%
\sim \omega ^{-\gamma /(\nu z)}$ in the scattering intensity for $\omega
\rightarrow 0$, while $\omega ^{\gamma /(\nu z)}\mathrm{Im}\chi \left(
\omega +i0^{+}\right) $ obeys $\omega /T$-scaling.

In the remaining paragraphs we discuss experiments, simulations and  the
generality of our results.
The NLSM (\ref{eq:action}) is the exact low-energy
theory of a dimer-diluted bilayer Heisenberg QAFM \cite%
{bilayerS,bilayerV,us_hsb04}, or in $d>2$, that of an appropriately \emph{%
bond}-diluted system. One candidate material in 3d is (Tl,K)CuCl$_3$ \cite{tlkcucl};
interesting quasi-2d compounds are SrCu$_2$(BO$_3$)$_2$ or BaCuSi$_2$O$_6$,
where suitable dopants remain to be found. In addition, the action (\ref{eq:action})
describes Josephson junction arrays with random missing junctions and diluted bosons in
optical lattices.

If magnetic ions are replaced by nonmagnetic ones in a
QAFM, e.g., Cu by Zn or Mg in La$_{2}$CuO$_{4}$ or YBa$_{2}$Cu$%
_{3}$O$_{6}$, impurity induced moments arise which leads to random Berry phases in the
NLSM \cite{SachdevVojta}. These moments produce Curie type contributions that
dominate the low-temperature behavior in the disordered phase $p>p_{c}$. However, at
$p=p_c$, their contributions are subleading compared to the critical singularities.
Thus, our theory also describes the percolation quantum phase transition
in site-diluted systems,
and we propose to study  QAFMs such as La$_{2}$Cu%
$_{1-p}$(Zn,Mg)$_{p}$O$_{4}$.
% using, e.g., inelastic neutron scattering.
Recent measurements \cite{Vajk_exp} indeed
gave a nontrivial dynamical exponent (with a somewhat smaller value of $%
z\approx 1.4$). The measurement of several observables may
help to identify the scaling regime and possibly refine the value of $z$. An issue that
may limit the applicability of our theory very close to $p_{c}$ is dilution induced
frustration.

We now turn to simulations, focusing on the dynamical exponent $z$. Early series
expansions
%of a square lattice Heisenberg model
\cite{Wan} gave values in the range of $1.5$ to $2$. Kato et al. \cite{Kato} obtained
spin-dependent values between $1.3$ and $2.5$ but under the assumption that the critical
percolation cluster is not long-range ordered. Recent simulations of a bilayer QAFM
\cite{bilayerS} and of a bond-diluted 2d QAFM \cite{Haas} gave values of about $1.9$, in
excellent agreement with our theory. These two papers also gave arguments for $z=D_{f}$
as obtained in our theory for $z_{0}=1$.

Finally, we discuss the generality of our theory. To derive the nonclassical scaling
theory (\ref{eq:scaling}) of the percolation quantum phase transition we only relied on
the fact that the energy gap $\omega _{0}$ of a cluster depends on its size $s$ via a
power law, $\omega _{0}\propto s^{-\varphi }$ with $\varphi
>0$.
%If this is fulfilled,the transition displays power-law scaling with nonclassical exponents.
If $\varphi$ diverges, which happens in our model for $z_0 = 2$, the relation between the
gap and $s$ becomes exponential. The transition is then of infinite-randomness type as in
the diluted transverse field Ising model \cite{SenthilSachdev}.  These observations agree
with a general classification of dirty phase transitions according to the effective
dimensionality of the droplets or clusters \cite{Vojta05}. For $z_{0}<2$, the clusters
(which are finite in space but infinite in imaginary time) are below the lower critical
dimension $d_c^-$ of the problem resulting in power-law scaling (and exponentially weak
Griffiths effects). In contrast, for $z_{0}=2$, the clusters are exactly at $d_c^-$. This
leads to activated scaling and power-law  Griffiths effects.

We acknowledge helpful discussions with M.\ Greven, M.L.\ Lyra, A.\ Sandvik, J.\ Toner,
and M.\ Vojta. Parts of this work have been performed at the Aspen Center for Physics and
the Kavli Institute for Theoretical Physics, Santa Barbara. This work was supported in
part by Ames Laboratory, operated for the U.S. Department of Energy by Iowa State
University under Contract No. W-7405-Eng-82 (J.S.), by the NSF under grant Nos.
DMR-0339147 (T.V.) and PHY99-07949 (KITP SB) as well as by
Research Corporation.


\begin{thebibliography}{99}
\bibitem{ramification} For Ising symmetry, this appears counterintuitive as
the dimension of the critical percolation cluster is larger than 1. However, for
fractals, topological factors like ramification are important for magnetic ordering
\cite{Gefen}.
\bibitem{subscript} In the following, classical percolation exponents carry a subscript $c$ while the magnetic exponents at our percolation
quantum phase transition do not carry a subscript.
\bibitem{fluc} For simple quantum antiferromagnets, $g$ can be related to the spin value
       $S$. In the large-$S$ limit, $g \sim 1/S^2$ \cite{sachdevbook}.
\bibitem{exact} This analysis gives the leading behavior exactly as
                the stable fixed point does not have anomalous dimensions.
\end{thebibliography}
\end{document}